\documentclass[a4paper, 10 pt, conference, dvipsnames]{ieeeconf}  % Comment this line out if you need a4paper

   \makeatletter
   \let\NAT@parse\undefined
   \makeatother

\IEEEoverridecommandlockouts                              % This command is only
\usepackage[%
  hyperindex=true, %
  colorlinks=true,%
  pagebackref=false,%
  plainpages=false,%
  pdfpagelabels,%
  linkcolor=NavyBlue,%
  citecolor=NavyBlue,%
  filecolor=NavyBlue,%
  urlcolor=NavyBlue%
]{hyperref}
\usepackage[disable,textsize=footnotesize]{todonotes}
\usepackage{lipsum}
\usepackage{placeins}
\usepackage[tracking=false,kerning=true,spacing=true]{microtype}
\usepackage{mdwlist}

\let\itemize\undefined
\makecompactlist{itemize}{stditemize}

\let\enumerate\undefined
\makecompactlist{enumerate}{stdenumerate}

\usepackage{cite}
\usepackage{graphicx} % for pdf, bitmapped graphics files
\usepackage{amsmath} % assumes amsmath package installed
\usepackage{amssymb}  % assumes amsmath package installed

\usepackage{amsthm}  
\theoremstyle{plain}
\newtheorem{assumption}{Assumption}
\newtheorem{proposition}{Proposition}
\newtheorem{definition}{Definition}

\theoremstyle{remark}
\newtheorem{remark}{Remark}
\usepackage{enumerate}
\usepackage{subfigure}

\usepackage{varioref}

\usepackage[tracking=false,kerning=true,spacing=true]{microtype}

\title{\LARGE \bf
\emph{Today Me, Tomorrow Thee}: Efficient Resource Allocation\\in Competitive Settings using Karma Games
}

\author{Andrea Censi, Saverio Bolognani, Julian G. Zilly, Shima Sadat Mousavi, Emilio Frazzoli % <-this % stops a space
\thanks{ The authors are with the Institute for Dynamic Systems and Control~(IDSC) and the Automatic Control Laboratory (IfA) at ETH Zurich, 8092 Zurich, Switzerland. E-mail:
        {\footnotesize \{acensi, bsaverio, jzilly, mousavis, efrazzoli\}@ethz.ch}
}%}}%
\thanks{\textcopyright 2019 IEEE.  Personal use of this material is permitted.  Permission from IEEE must be obtained for all other uses, in any current or future media, including reprinting/republishing this material for advertising or promotional purposes, creating new collective works, for resale or redistribution to servers or lists, or reuse of any copyrighted component of this work in other works.}%
}

\newcommand{\expectation}[1]{\mathbb{E}\left[#1\right]}

\newcommand{\PK}{\mathcal{K}}
\newcommand{\PM}{\mathcal{M}}
\newcommand{\PO}{\mathcal{O}}
\newcommand{\PD}{\mathcal{P}}
\newcommand{\maxcarma}{k_{\max}}

\newcommand{\ft}{\phi}
\newcommand{\po}{\gamma}
\newcommand{\PS}{\mathcal{U}}

\newcommand{\realp}{\mathbb{R}_{+}}

\newcommand{\transitions}{T}
\newcommand{\stationary}{D}
\newcommand{\policy}{\pi}
\newcommand{\avgcost}{\overline{c}}
\newcommand{\exputility}{\theta}

\DeclareMathOperator{\variance}{var}

\usepackage[absolute,overlay]{textpos}

\hypersetup{pdftitle={Today Me, Tomorrow Thee: Efficient Resource Allocation in Competitive Settings using Karma Games},pdfauthor={Andrea Censi, Saverio Bolognani, Julian Zilly, Shima Sadat Mousavi, Emilio Frazzoli}}

\begin{document}

\begin{textblock*}{\textwidth}(15mm,18mm) % {block width} (coords) 
\centering \bf \textcolor{NavyBlue}{To appear in the Proceeding of the 22nd IEEE Intelligent Transportation Systems Conference, ITSC 2019.}
\end{textblock*}

\maketitle

% \begin{textblock*}{\textwidth}(15mm,65mm) % {block width} (coords) 
% \begin{center}\bf \textcolor{NavyBlue}{Published on \emph{IEEE Transactions on Power Systems,} vol. 33, no. 6, pp. 6705-6714, Nov. 2018.}\end{center}
% \\
% \begin{center}\url{http://doi.org/10.1109/TPWRS.2018.2850448}\end{center}
% \end{textblock*}

\thispagestyle{empty}
\pagestyle{empty}

%%%%%%%%%%%%%%%%%%%%%%%%%%%%%%%%%%%%%%%%%%%%%%%%%%%%%%%%%%%%%%%%%%%%%%%%%%%%%%%%
\begin{abstract}
We present a new type of coordination mechanism among multiple agents
for the allocation of a finite resource, such as the allocation of time slots for passing an intersection.
We consider the setting where we associate one counter to each agent, 
which we call \emph{karma value}, and where there is 
an established mechanism to decide resource allocation based on agents exchanging karma.
The idea is that agents might be inclined to pass on using resources today, in exchange for karma, which will make it easier for them to claim the resource use in the future. 
To understand whether such a system might work robustly, we only design the protocol and not the agents' policies.
We take a game-theoretic perspective and compute policies corresponding to Nash equilibria for the game.
We find, surprisingly, that the Nash equilibria for a society of self-interested agents are
very close in social welfare to a centralized cooperative solution.
These results suggest that many resource allocation problems can have a simple, elegant, and robust solution, assuming the availability of a karma accounting mechanism.
\end{abstract}

%%%%%%%%%%%%%%%%%%%%%%%%%%%%%%%%%%%%%%%%%%%%%%%%%%%%%%%%%%%%%%%%%%%%%%%%%%%%%%%%
\section{Introduction}

The very survival and success of a society with shared resources depends 
on the rules and protocols agents use to interact with each other. 

In designing the rules of these societies, there is always a trade-off concerning
centralization, efficiency, robustness, and resiliency. A centralized system for resource allocation
needs more infrastructure and is less robust and resilient, yet it is the most efficient.
A distributed system is more resilient and privacy-preserving.

In intelligent transportation systems, we can distinguish the ``macro'' level of the fleet, and the ``micro'' level
of the vehicles. 
At the macro level, much research has shown how it is possible to obtain a substantial improvement in the efficiency of a transportation network~\cite{samaranayake2017ridepooling, ruch2018amodeus} by optimizing resource use through cooperative approaches; that is, one takes the perspective of a single agent which is able to control centrally a fleet of vehicles.
At the micro level there are similar resource allocation problems. Because of the advent of self-driving cars to be used in autonomous mobility on demand networks, the `micro' coordination problems become interesting, as we study how the codes, customs, and conventions of human drivers can be generalized to a scenario with both artificial and human agents.

\begin{figure}[ht]
    \centering
    \includegraphics[height=3.5cm]{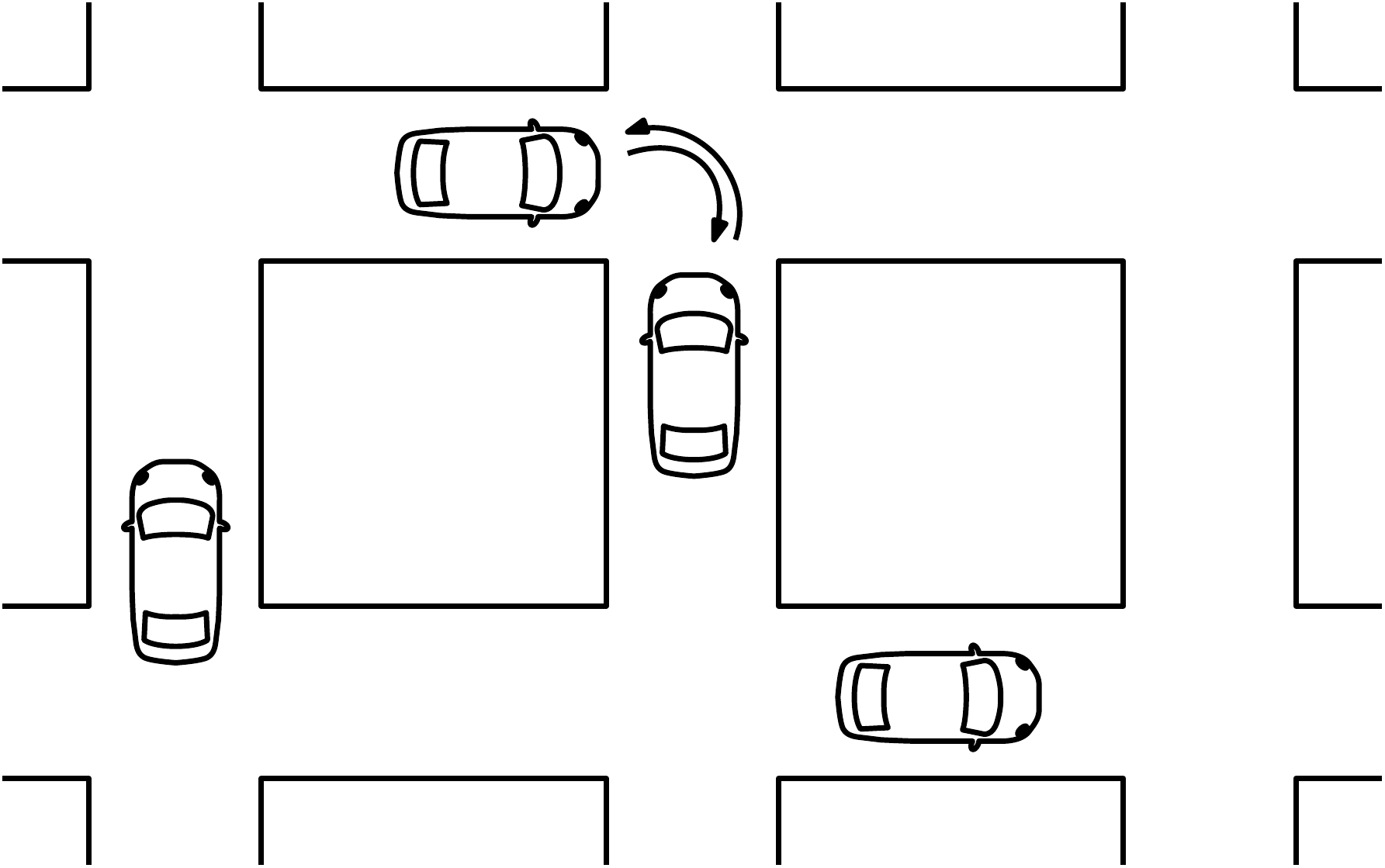}
    \caption{
    We propose an innovative approach to the problem
    of resource allocation in a competitive setting based
    on the notion of ``karma'', an accounting system that summarizes
    the agent's actions in the past. % as well as its politeness and luck. 
    The karma system allows agents to accept to give in at a particular 
    interaction, while receiving a karma compensation. This allows an overall more efficient use of resources. Agents interact by meeting in pairs, e.g. $\{i, j\}$, and bid on the resource by sending messages $\{m_{i}, m_{j}\}$ specifying how much karma they are willing to bid in that particular interaction. The agent with the larger bid wins and gets access to the limited resource which in this case is access to an intersection leading to no delay $\delta$ in travel time for the winner.
    }
    \label{fig:interaction_oerview}
\end{figure}
The prototypical problem is intersection management. Deciding which car may pass first is a resource allocation problem, in which the resource is the use of the space inside the intersection in a given time interval. Similar resource allocation problems happen also in maneuvers outside of intersections, as drivers compete for the use of space, although the outcome is not as simple as a discrete decision as in intersection management. These interactions happen between independent agents, with competitive goals, and typically are not repeated, as it is rare to encounter the same vehicle again. Therefore, there is little incentive to give in at one interaction; at face value, this appear to be a non-repeated game. 

Typical human drivers do not act like self-interested agents. Humans have ways to communicate urgency and politely negotiate maneuvers while they drive. 
Ultimately this is due to the altruism and pro-sociality bias that evolved in our species~\cite{doi:10.1146/annurev-psych-010814-015355}; the bias makes the single individual intrinsically happy to accommodate somebody who seems to be in a hurry. Our species thrived because individuals are not completely self-interested. When we lived in tribes, deviant antisocial behavior was easily spotted and repressed; now that our social groups are counted in the billions, a set of rules (laws) and corresponding incentives (punishments) help in aligning the individual and societal interests in the handling of common resources~\cite{ostrom2015governing}. When driving, some of our behaviors derive from these incentives (we do not speed because we are afraid of tickets), but many polite behaviors are due to our visceral intrinsic motivation rather than extrinsic rewards/punishment.

How can we ensure that a population of artificial agents, such as self-driving cars, can attain the same efficiency of a pro-social species like humans?  In this paper, we consider the problem of resource allocation in a setting that we call \emph{Karma Game}.
The idea is that considerable gains can be realized if an agent is inclined to give in at one interaction,
if it is compensated with ``karma''. Thus, we introduce karma as a way to account for an agent's past actions.
(This concept is closer to how ``karma'' is used in video game mechanics, 
rather than to how it is understood in Indian religions.)

We define a karma protocol with which agents can negotiate the use of resources. The protocol 
describes the exchange of bidding messages and how karma is updated based on the outcome of the interaction. 
The protocol does not need a third party, and the primitives needed
to implement karma accounting and the interaction are those provided by many blockchains, such as Ethereum~\cite{wood2014ethereum}.

Having fixed the protocol, we study how a population of self-interested agents will use it, by computing 
the Nash equilibria for the resulting Karma game.  We then compare the Nash equilibria of the distributed system with the baseline of the optimal centralized policies. We observe that the efficiency of the system is remarkably similar. The social welfare is thus closely aligned with the self-interest of the agents, assuming the agents have reasonable discount factors. An agent that does not
care about the future and lives for the present will also create an inefficient society.

\section{Related work}

\paragraph{Intersection control}

Traditional intersection control strategies have been substantially based on utilizing control devices such as traffic lights, in which an offline optimization based on historical data can be used to provide a control signal~\cite{robertson1969transyt,sims1980sydney}. The main  drawback of this control strategy is that it cannot adapt to changes in request patterns and environment. 
Improving upon classical control strategies, communication-based schemes~\cite{chen2016cooperative} are based on a competitive scenario, in which different vehicles aim at minimizing their own selfish cost.
It is  assumed that the urgency~$u_i(t)$ is a piece of \emph{private information} of each vehicle~$i$, and is therefore not accessible to other vehicles. This kind of scenario is typically tackled via auctions, which can be designed in order to induce selfish agents to disclose their true urgency~\cite{schepperle2007towards,carlino2013auction,
levin2015intersection,vasirani2012market,mashayekhi2015multi,
raphael2017intersection, isukapati2017accommodating, sayin2018information}. 
For example, in \cite{schepperle2007towards}, the earliest time-slot in an intersection is auctioned off by an intersection manager among all vehicles at the front  of each lane. \todo[inline]{the following sentence is unclear}In \cite{carlino2013auction}, having an infinite budget, any agent in a lane  can participate in a second-price auction to enhance the winning chance of the agent at the front. In  \cite{levin2015intersection}, a mechanism based on a first-price auction is proposed for the management of intersections.  Two scenarios for  single intersection and a network of intersections are considered in \cite{vasirani2012market}, and a policy based on a combinatorial auction for assigning the reservations of time-space slots is presented. However, finding the winner of a combinatorial auction is NP-hard~\cite{nisan2007algorithmic}. Finally, to schedule the intersection usage, \cite{sayin2018information} proposes a variant of the Vickrey-Clarke-Groove mechanism in which an intersection unit charges each agent at the front of any lane with a time-token based on its impact on others.

We note that our approach departs from the auction-based schemes in the mentioned papers in that to maintain the fairness properties between wealthier drivers and those without many funds, it does not require any monetary transactions, and therefore does not require to attach an objective value to the cost incurred by the vehicles.
We will discuss later how this sheds light on the true nature of this coordination problem. Any vehicle is assigned an initial karma level. In light of the budget-balance property of our mechanism, the total amount of karma remains constant over the whole transportation network. Also, unlike the assumption in \cite{carlino2013auction}, every agent is  assumed to have a \emph{limited} total karma at any time period, which neither is negative, nor exceeds a maximum value.

  Almost all works in the literature which proposed an auction-based approach for the intersection control are static, one-time decision problems. However since the urgencies and the agent's private information 
change over time, a sequence of decisions needs to be made, resulting in a dynamic resource allocation and a dynamic bidding process~\cite{bergemann2010dynamic}. Thus, the utility function of each agent along with the social welfare are defined based on the discounted utility over time. We assume that in every interaction, vehicles are allowed to communicate a scalar \emph{message} $m_i(t)$.
The karma value of each agent is a public state $k_i(t)$.
Both the outcome of the interaction (who goes first) and the update of the public state $k_i$ are determined based on a set of rules which are known and verifiable to all agents (as they only depend on public information: the states $k_i,k_j$ and the messages $m_i,m_j$). 

\paragraph{Karma-like concepts}

  A   ``karma'' system was   introduced in \cite{vishnumurthy2003karma} in the context of file-sharing  to prevent ``freeloading'' in peer-to-peer networks. In this framework, karma represents the standing of each agent in the system, that increases when contributing and decreases when consuming a resource, and thereby incentivizes agents to contribute resources~\cite{garcia2004off}. 
  In this and similar systems, the ``value'' of the karma is fixed---in our approach, the agents are free to assign a value to karma according to their goals and current state.
  
\paragraph{Population games}

This competitive scenario can be modeled as a repeated game (interactions) between randomly selected agents in a large population.
For the analysis of the resulting game, we adopt the approach that is typically used in the study of \emph{population games} \cite{Sandholm2010}, which has its underpinnings in the following abstractions:
1)~populations are continuous rather than discrete; the payoffs to a given strategy therefore depend on society's aggregate behavior in a continuous fashion;
2)~the aggregate behavior in a population game is described by a ``social state'', which specifies the empirical distribution of strategy choices (or \emph{types}) in the population; for simplicity, this social state is generally finite-dimensional.
The specific application that we are considering has however some peculiarities, compared to standard population games: for example, each agent's \emph{type} is also determined by an exogenous time-varying signal (their urgency). Moreover, there is no natural \emph{revision protocol} or \emph{adaptation}, and therefore no evolution of the agents. 
We therefore prefer to present the resulting game in a self-contained and specialized form, without explicitly tapping into that literature for definitions or results. 
Notice that the game we are formulating is more general than the specific traffic interaction problem, although clearly inspired by that setup.         

\section[Resource allocation in a drive-by scenario]{Resource allocation in a ``drive-by'' scenario}
\label{section:model}

In this section we introduce a deliberately simple model for vehicle-to-vehicle interaction at intersections. 
We strove to simplify the model to its core features, in order to isolate the essential phenomena in this problem.
We understand the problem of vehicle-to-vehicle interaction at intersections as an example
of a ``drive-by'' scenario, in which:
\begin{enumerate}
    \item There is a large number of agents in the systems.
    \item Agents interact with a random schedule.
    \item Each agent interacts many times with other agents over its lifetime.
    \item The value of a resource to an agent varies in time according to an exogenous factor.
\end{enumerate}

For vehicle-to-vehicle (V2V) interactions at intersection: 
\begin{enumerate}
    \item There is a large number of cars on the road.
    \item Cars meet randomly at intersections.
    \item Each car encounters many intersections over its lifetime.
    \item The value of time saved to a car varies in time according to its urgency
    on that day.
\end{enumerate}

\subsection{Formalization}

More formally, consider a population $\mathcal N$ of $N$ vehicles.
Each vehicle $i \in \mathcal N$ has an associated \emph{urgency} process $u_i(t)$. The urgency $u_i$ at time $t$ indicates the marginal value that agent $i$ gives to a unitary delay in its trip. It is an exogenous process that is not affected by the behavior of the vehicles.

The vehicles interact at intersections. Each interaction at time $t$ involves only a 
pair of vehicles $\mathcal I(t) =\{ i, j\} \subset \mathcal N$.

Every time two vehicles interact, one of the two vehicles is necessarily delayed by a unitary delay, 
while the other vehicle does not incur any delay.
We therefore have two possible outcomes $o(t)$, that is~$o(t) \in \mathcal O := \{i,j\}$.
Agent~$i$ (and, in a completely symmetric way, agent~$j$) incurs a cost~$c: \PO \times \PS \to \realp$
that is a function of the outcome and of its own urgency, and is defined as
\begin{equation}
c_i(o,u_i) =
\begin{cases}
u_i, & \text{if }o=i\text{;}\\
0, & \text{otherwise.}
\end{cases}
\label{eq:costfunction}
\end{equation}

\subsection{Assumptions}

We propose the following assumptions about the model. 

\begin{assumption}[Randomness of encounters]
\label{assumption:iidinteraction}
The sequence $\mathcal I(t)$ is random and identically distributed at all times $t$ over the set $\{\mathcal I \subset \mathcal N,  |\mathcal I| = 2\}$, and each vehicle has the same probability of belonging to $\mathcal I(t)$ at a given $t$.
\end{assumption}

\begin{assumption}
\label{assumption:equalurgency}
The urgency processes $u_i(t)$ are identical for all vehicles $i\in\mathcal N$.
The urgency at each time $t$ is independent and identically distributed, and takes values in $\mathcal U := \{0, U\}$.
\end{assumption}

\todo[inline]{I made it $0$ and $U$ instead. I don't have a strong preference, but I guess urgencies need to be ordered, and this is not apparent from the labels.}

We defer the discussion on how to relax these assumptions to Section~\ref{section:conclusions}.
For the most part, these assumptions are introduced for technical convenience, as they yield a simpler analysis, a computational advantage (see also Section~\ref{section:computing}), and a more immediate interpretation of the results.

\subsection{Performance measures}

\label{section:social}

The focus of this paper is on \emph{policies} that allow to decide $o(t)$ optimally, where the notion of optimality is to be defined hereafter.

We define two measures of social cost for the entire population, which are associated to two different interpretations.
The first measure simply quantifies the expected aggregate cost for the entire system at each interaction:
\[
W_1 := \mathbb{E}\big[\, \textstyle{\sum_{\ell\in \mathcal N}}\  c_\ell(o(t),u_\ell(t))\, \big].
\]
The second measure quantifies the expected rate at which the variance (across agents) of the accumulated cost grows:
\[
W_2 = \lim_{t \rightarrow \infty}
\expectation{\variance a(t+1) - \variance a(t)}
\]
where 
\[
\variance a := \frac{1}{N} \textstyle{\sum_{\ell \in \mathcal N}}
\left(
a_\ell - \frac{1}{N} \textstyle{\sum_{k \in \mathcal N}}\, a_k
\right)^2
\]
and $a$ denotes the vector of accumulated costs of the agents, defined element-wise as
$
a_\ell(t) = \sum_{\tau=0}^t c_\ell(o(t), u_\ell(t)).
$
In these expressions, $\expectation{\cdot}$ represents the expectation with respect to both the stochastic urgency processes and the interaction selection process (which are independent processes).

\subsection{Centralized policies}

In this section, we derive the optimal centralized policies for the simplified intersection management problem that we presented, under the notions of social optimality that we described. 
These optimal centralized policies will constitute a baseline for the analysis of the policies that emerge in a distributed competitive setting.

In a centralized setting, we are allowed to adopt causal policies of the kind
\[
o(t) = \Pi
\left(
\mathcal I(t), \left\{u(\tau)\right\}_{\tau=0}^{t},  \left\{o(\tau)\right\}_{\tau=0}^{t-1}
\right),
\]
where by $u(t)$, we indicate the past urgencies of all agents.

Under Assumptions~\ref{assumption:iidinteraction} and \ref{assumption:equalurgency}, 
the optimal policies for the two social costs $W_1$ and $W_2$ can be computed explicitly.

\todo[inline]{ref to these two equations in the numerical experiments}
\begin{proposition}
\label{proposition:optimalpolicies}
The social costs $W_1$ and $W_2$ are minimized, respectively, by the policies
\begin{equation}
o_1^*(t) \in \arg\min_{\ell \in \mathcal I(t)} u_\ell(t) \label{eq:centralized-urgency}
%\label{eq:ow1}
\end{equation}
and
\begin{equation}
o_2^*(t) \in \arg\min_{\ell \in \mathcal I(t)} a_\ell(t-1) + u_\ell(t). \label{eq:centralized-cost}
%\label{eq:owinfty}
\end{equation}
\end{proposition}
\todo[inline]{Should we include the proof? It's trivial for the first case, not for the second one (I honestly have not written it down, although simulations suggest that it's correct.}

If the $\arg \min$ operation does not return a singleton, then any of the two choices is optimal. 
Here and thereafter,   we assume that  $\arg \min$ ties are resolved via fair coin flipping.

We also define a third centralized policy, which prioritizes the minimization of $W_1$ (therefore obtaining the same value for $W_1$ as $o_1^*$) and, in case of ties between the urgencies $u_i$ and $u_j$ (where $\mathcal I = \{i,j\}$), aims at minimizing the unfairness defined by $W_2$:
\begin{equation}
\begin{split}
o_{1,2}^*(t) \in &\arg\min_{\ell \in \mathcal I(t)} u_\ell(t)\\ &\text{and} \\
 u_i(t) = u_j(t) \Rightarrow
 o_{1,2}^*(t) \in &\arg\min_{\ell \in \mathcal I(t)} a_\ell(t-1) + u_\ell(t).
\end{split}
\label{eq:centralized-urgency-then-cost} %\label{eq:oonetwo}
\end{equation}

\section{Resource allocation using Karma Games}

In this section, we formulate a mechanism for resource allocation based on the notion of karma.
We only design the mechanism and not the agents' policy, which is going to be found automatically through optimization.

\subsection{Informal definition of karma interaction mechanism}

We assume that there is an integer counter $k_i(t)$ (karma) associated to each agent bounded by $\maxcarma$.
The agents exchange one message  at each interaction. Each agent $i$ can produce 
a message $m_i$ which contains a value not to exceed its current karma:\quad$
0 \leq m_i(t) \leq k_i(t).
$

We give this message the semantics of how much karma the agent sees fit to bid on the current interaction.
The agent that provides the highest message is allowed to use the resource (go first at the intersection) and 
must pay the other agent \emph{up to} the karma value that it has bet. 
The karma transferred is reduced if the transfer would make the 
other agent overflow~$\maxcarma$. Suppose that agent~$i$ wins betting~$m_i$.
Then the karma transferred is $\min(m_i, \maxcarma - k_j)$.

\begin{remark}
In this paper we do not delve into the technical implementation of such a scheme, but we would like
to remark that it is possible to implement such a scheme, in a completely distributed way,
without an arbiter to preside at each interaction, by using some of 
the cryptographic primitives associated to blockchain technology. The counters are implemented using public addresses.
Non-refutable messages are implemented using cryptographic commitments. The resolution and the outcome 
can be easily implemented using the primitives of, for example, Ethereum's Solidity language.
\end{remark}

\subsection{Formal definition of Karma Game}

We formalize the discussion so far by defining \emph{Karma Games} in a way that is 
slightly more general.

\begin{definition}[Karma Game in Tabular Format]\label{def:karma-game}
A Karma Game $G$   is a tuple 
\[
G = \langle \PK, \PM, \PO, \PS, p, c, \alpha, \po, \ft \rangle,
\]
where:
\begin{itemize}
    \item $\PK $ is a set of possible public states (\emph{karma}) of an agent;
    \item $\PM $ is a set of possible \emph{messages} of an agent;
    \item $\PO $ is a  set of possible \emph{outcomes} of an interaction;
    \item $\PS$ is a set of possible \emph{exogenous states} of an agent and $p$ is a probability distribution on~$\PS$;
    \item $c: \PO \times \PS \to \realp$ is the instantaneous cost for each agent, which depends on the outcome of the interaction and on the exogenous state of the agent;
    \item $0\le\alpha<1$ is a discount factor;
    \item $\po: \PK \times \PM \times \PK \times \PM \to \PD(\PO)$ is the \emph{interaction outcome function}, as a probability distribution on~$\PO$;
    \item $\ft: \PK \times \PM \times \PK \times \PM \times \PO \to \PD(\PK)$ is the \emph{state transition function}.
\end{itemize}
The interpretation is as follows.
Suppose an agent of karma~$k_i(t)$ meets an agent
of karma~$k_j(t)$, and they exchange messages~$m_i(t)$ and~$m_j(t)$.
The function $\po$ gives a distribution on the possible outcome $o(t) \in \PO$
given by  
\[
o(t) \sim \po(k_i(t), m_i(t), k_j(t), m_j(t)).
\]
As for the consequences, $\ft$ is the map that specifies the probability distribution of the next value of~$k_i$ and $k_j$:
\[
k_i(t+1) \sim \ft(k_i(t), m_i(t), k_j(t), m_j(t), o(t) ).
\]
The cost for each agent is given by the following series,
where time is to be interpreted as ranging over the instants
in which the agent participated in an interaction:
\begin{equation}
    C = \mathbb{E} \Big[\,
    \textstyle{\sum_{t=0}}\, \alpha^t c(o(t), u(t)) 
    \,\Big].
\label{eq:totalcost}
\end{equation}
\end{definition}
\newcommand{\ifirst}{\text{``$i$ 1st''}}
\newcommand{\isecond}{\text{``$i$ 2nd''}}

\subsection{Vehicle interaction as a Karma Game}

We now put in the form of Definition~\ref{def:karma-game}, the model we described so far.
$\PK = \PM$ is the set of integers up to~$\maxcarma$:
$$
\PK = \{0, 1, 2, \dots, \maxcarma \}.
$$
There are two possible outcomes of each interaction, as explained in Section~\ref{section:model}: $\PO = \{ i, j \}.$
For the outcome distribution $\po(k_i, m_i, k_j, m_j)$, we have
\[
    \mathbb P(o = i) =  
    \begin{cases}
    0, & \text{if $\tilde{m}_i  > \tilde{m}_j $}, \\
    1, & \text{if $\tilde{m}_i  < \tilde{m}_j $}, \\
    0.5, & \text{if $\tilde{m}_i  = \tilde{m}_j $}, \\
    \end{cases}
\]
where we defined $\tilde{m}_i = \min(m_i, k_i)$.

For the state transition function  $\ft(k_i , m_i , k_j , m_j , o)$, we have that with probability 1 
\[
k_i(t+1) = 
\begin{cases}
k_i  - \min(\tilde{m}_i, \maxcarma - k_j),& \text{if } o(t)=i, \\
\min(k_i  + \tilde{m}_j, \maxcarma), & \text{if } o(t)=j.
\end{cases}
\]
These rules guarantee that 
\begin{itemize}
    \item the total amount of karma is conserved.
    \item karma is bounded above by $\maxcarma$ and below by $0$.
\end{itemize}

The cost function $c$ is the one already defined in \eqref{eq:costfunction}.

\section{Acting rationally in a Karma Game}

We now turn attention to what is the rational behavior of an agent in a Karma Game. 
An agent's behavior is completely defined by its policy. 

\begin{definition}[Agent policy]
In a Karma Game, the agent's policy $\policy$ is a probability distribution over the possible messages,
which varies as a function of the agent's current urgency $u_i(t)$ and current karma $k_i(t)$:
    \[
    m_i(t) \sim \policy(u_i(t), k_i(t)).
    \]
\end{definition}

As an agent, we need to decide what message to send for each combination of urgency $u_i\in\PS$
and karma $k_i\in\PK$. In game theory jargon, we speak of a set $\mathcal{A}$ of different agent ``types''; 
in this case, $\mathcal{A} \simeq \PS \times \PK$. The traditional notion of ``agent type'' does not fully
capture our setting; because following an interaction, the type of an agent changes as they gain/lose karma.
Moreover, the urgency is an exogenous variable that nobody can predict. Still, we use ``agent type'' in the following.

Under our assumptions, it is easy to compute the optimal policy for an agent if the urgency is zero. In that case,
the optimal action  for the agent is to send a message $m_i(t) = 0$. That is because the
agent is indifferent to losing or winning the interaction regarding the cost; and, regarding
the karma, the agent prefers to lose the interaction hoping to gain some karma.

If an agent has a nonzero urgency, how much karma should she bid today? This does not have an easy answer, except
in special cases. For example, if the discount factor $\alpha$ is zero---the agent does not care about 
the future, then the optimal policy is to send the maximum message $m_i(t) = k_i(t)$.
In all other cases, we need to characterize and compute Nash equilibria for this game.
Figure~\ref{fig:definitions-policy} shows a representation of such an optimal policy obtained as a Nash equilibrium.

\def\figureh{4.2cm}

\begin{figure*}
    \centering
    \subfigure[]{%
    \label{fig:definitions-policy}%
    \includegraphics[height=\figureh]{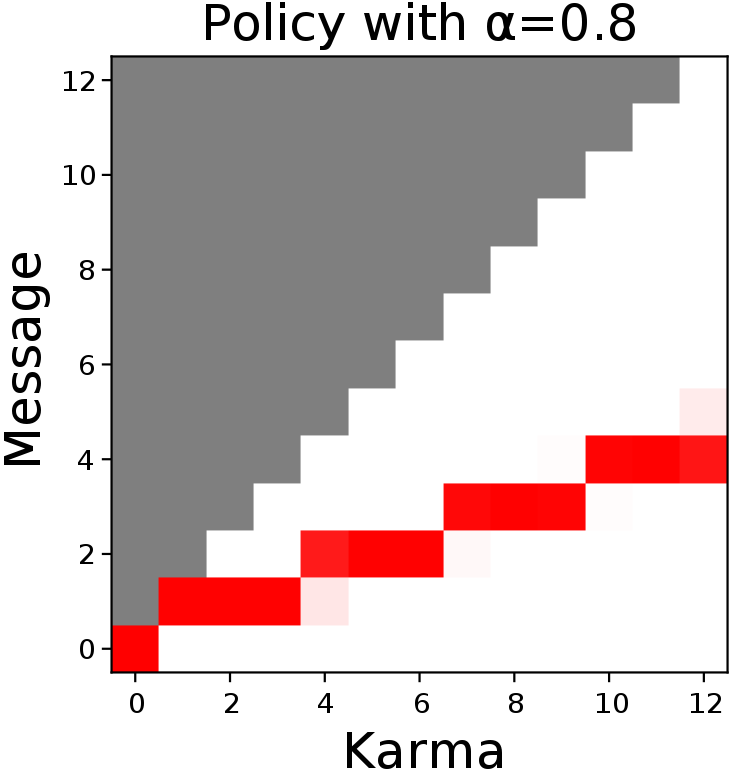}%
    }~ 
    \subfigure[]{%
    \label{fig:definitions-utilities}%
    \includegraphics[height=\figureh]{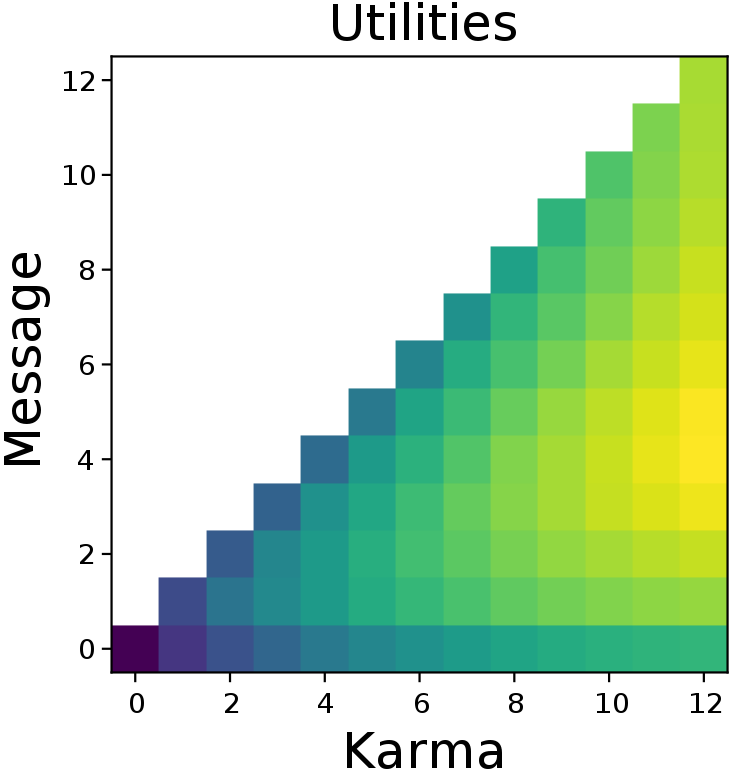}%
    }~ 
    \subfigure[]{%
    \label{fig:definitions-stationary}%
    \includegraphics[height=\figureh]{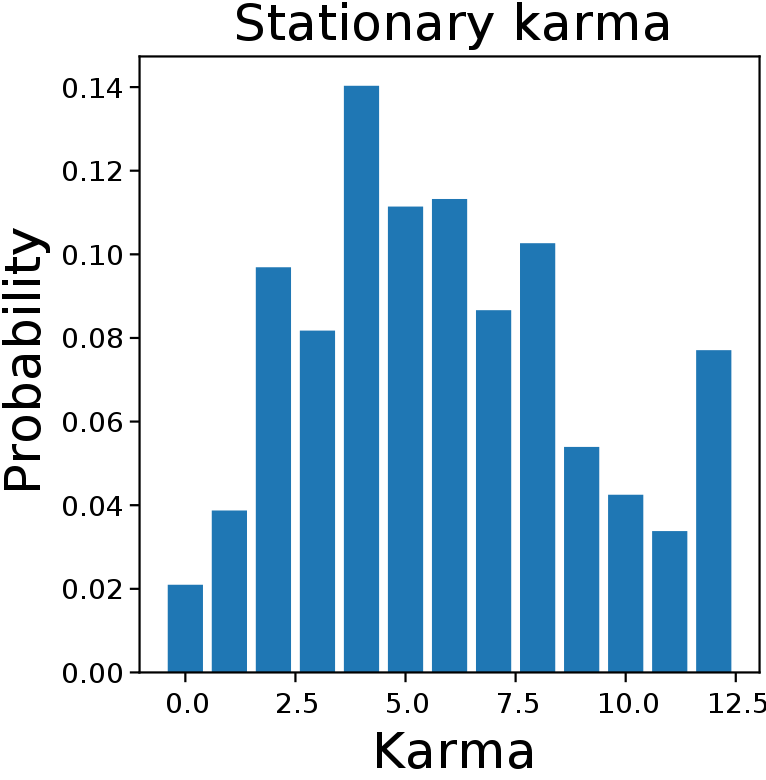}%
    }\\
    \subfigure[]{%
    \label{fig:definitions-final-utilities}%
    \includegraphics[height=\figureh]{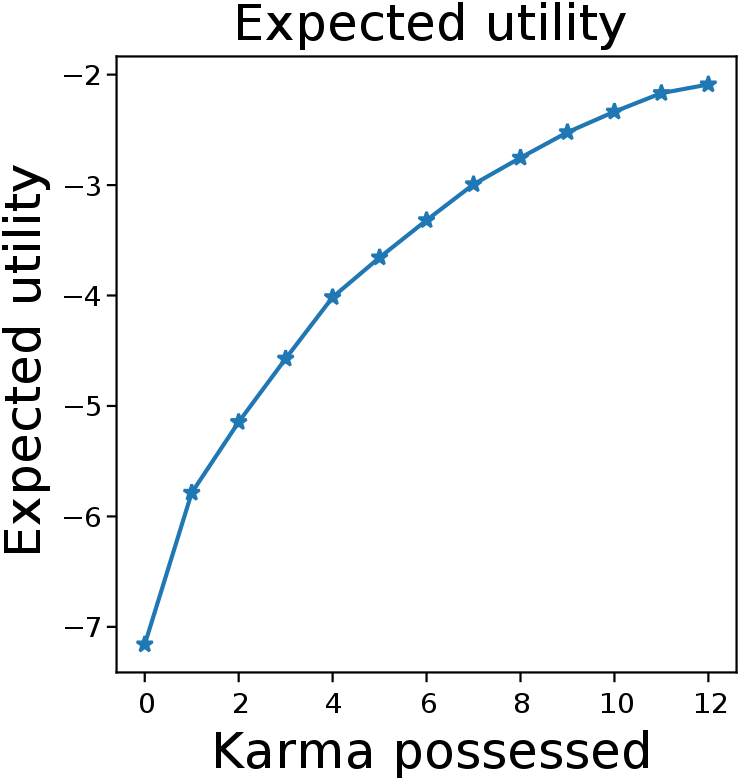}%
    }~ 
    \subfigure[]{%
    \label{fig:definitions-transitions}%
    \includegraphics[height=\figureh]{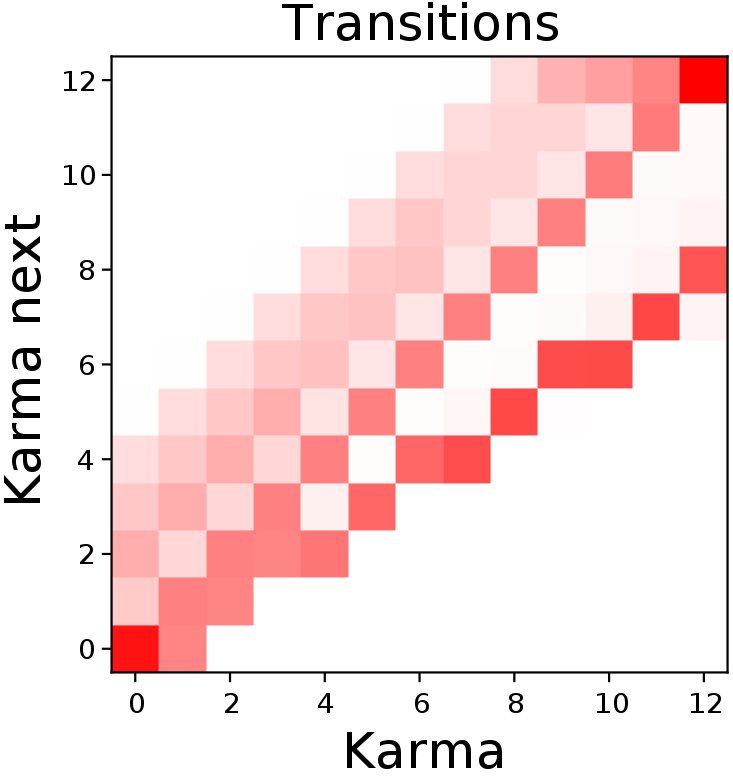}%
    }
\caption{For a Karma Game, the definition of a Nash equilibrium (Definition~\ref{def:nash_eq}) involves a policy, from which one can derive
the other quantities. (a): Optimal policy as mapping from current karma level to likelihood of sending a given message $m$; (b) The expected utility of sending each message $m$ as a function of karma levels.  (c) Stationary karma distribution across all agents; (d) The expected utility of possessing a certain karma level. (e) The transitions show the likelihood of transitioning from a certain karma level to another level.%either by bidding or not bidding karma. 
}
\label{fig:definitions}
\end{figure*}

\begin{figure*}
    \centering
    \subfigure[Iteration \#74]{%
    \includegraphics[height=\figureh]{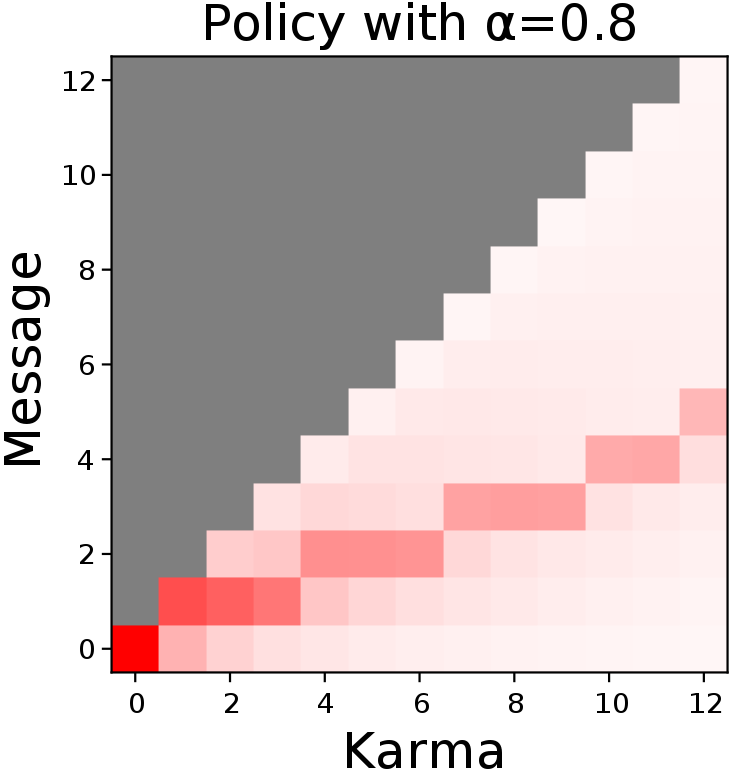}
    }
    \subfigure[Iteration \#305]{%
    \includegraphics[height=\figureh]{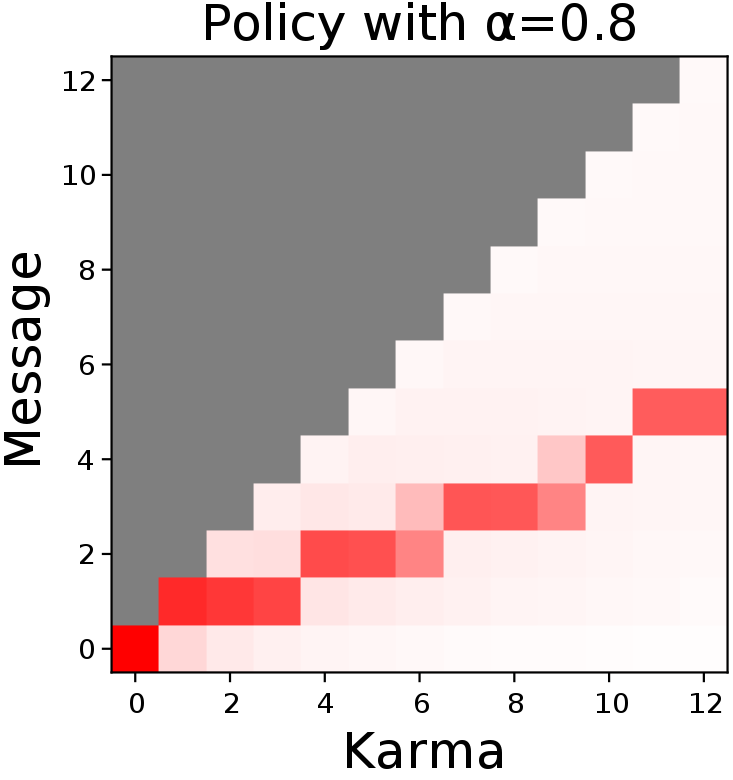}
    }
    \subfigure[Iteration \#705]{%
    \includegraphics[height=\figureh]{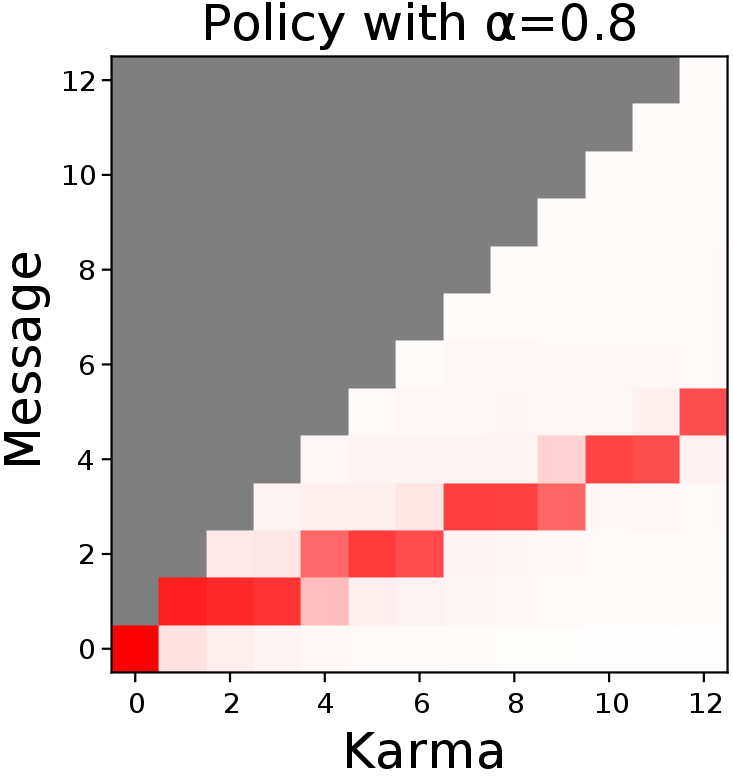}
    }
    \subfigure[Iteration \#954]{%
    \includegraphics[height=\figureh]{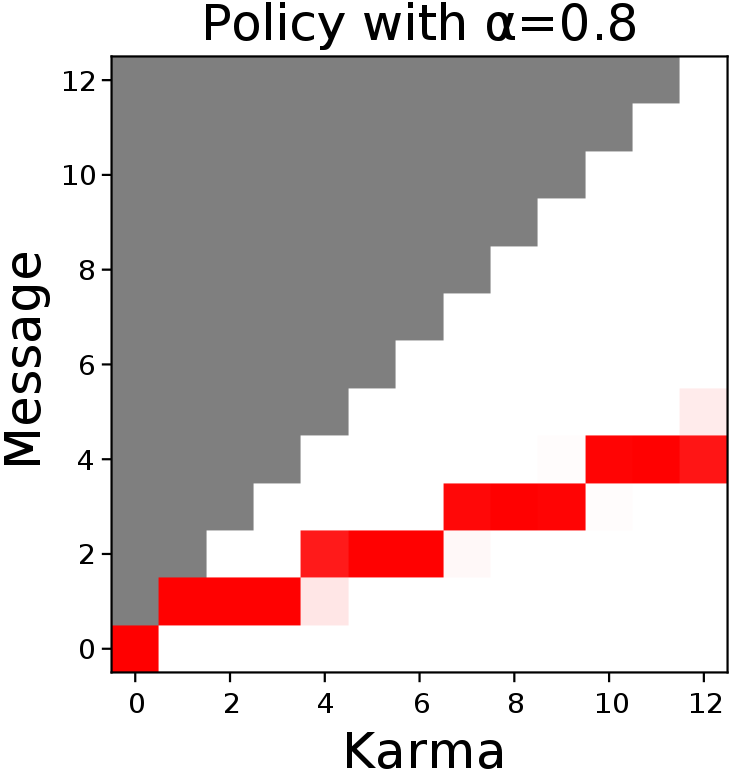}
    }
    \caption{Evolution of policy as the temperature decreases in the simulated annealing procedure. The policy becomes progressively more rational, until we find a Nash equilibrium.
    }
    \label{fig:policy_evolution}
\end{figure*}

\begin{figure*}
    \centering
    \subfigure[]{%
    \includegraphics[height=\figureh]{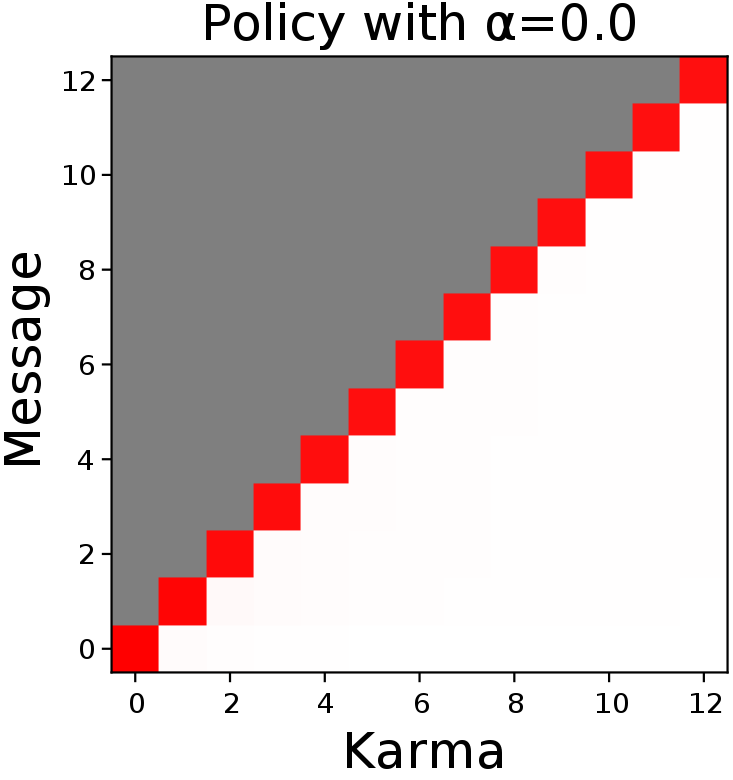}
    }
    \subfigure[]{%
    \includegraphics[height=\figureh]{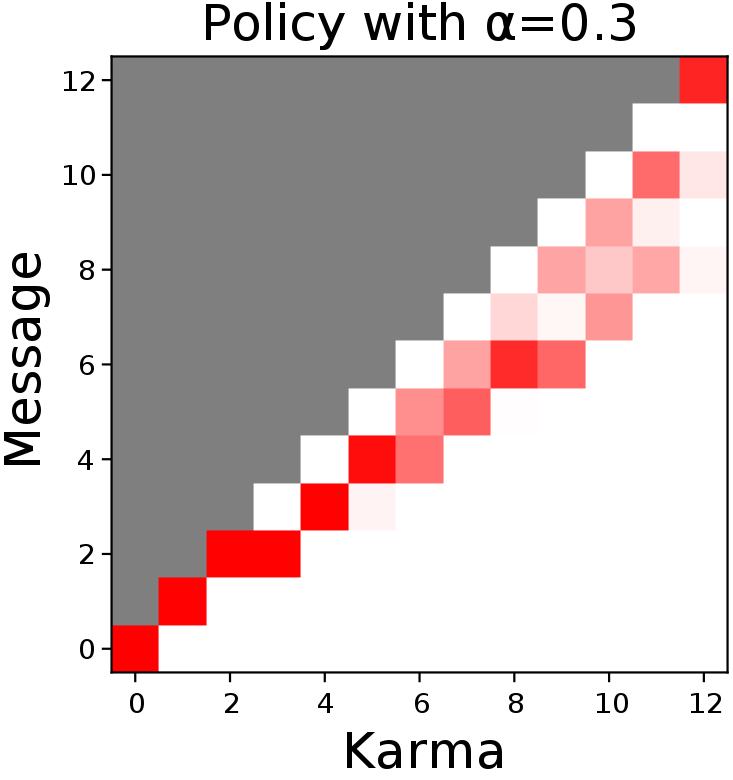}
    }
    \subfigure[]{%
    \includegraphics[height=\figureh]{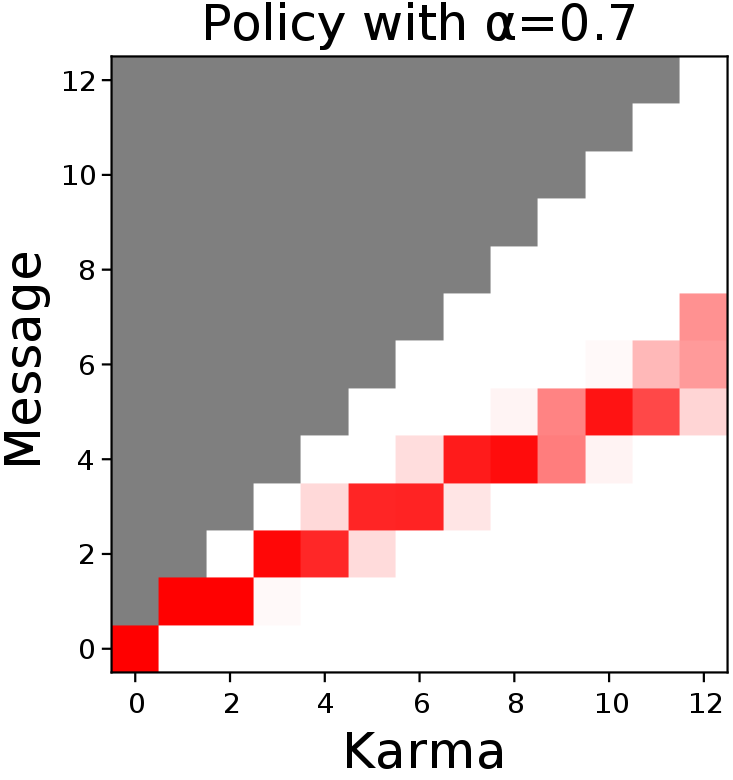}
    }
    \subfigure[]{%
    \includegraphics[height=\figureh]{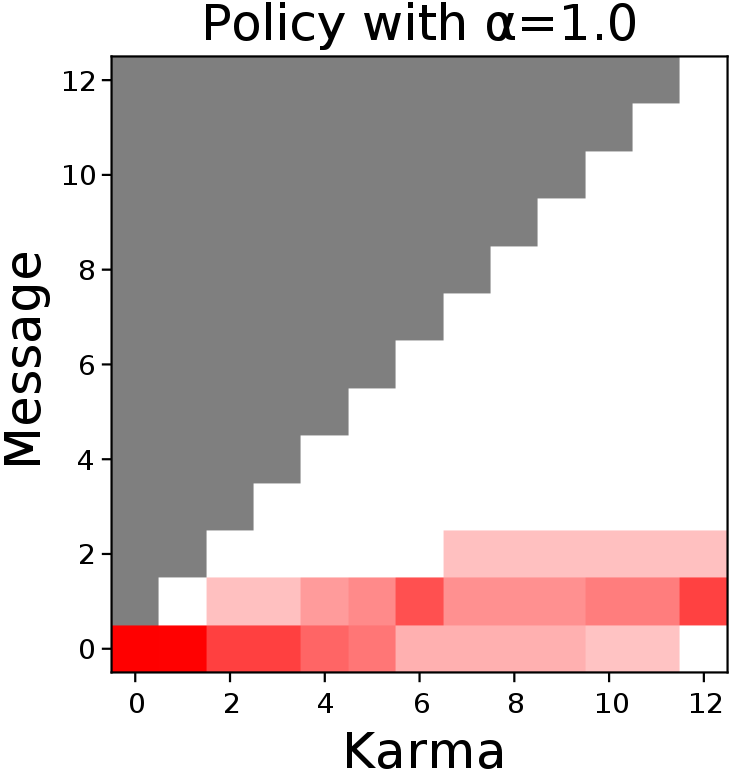}
    }
    \caption{We investigate the effect of discounting future rewards. Displayed are discounting values $\alpha$ in the set $\{0, 0.3, 0.7, 1.0\}$. As time discounting $\alpha$ increases from $0$ to $1$, the future is valued higher and higher and less karma is bid for the same karma levels.
    }
    \label{fig:discounting}
\end{figure*}

\begin{figure}
    \centering
    \includegraphics[width=\columnwidth]{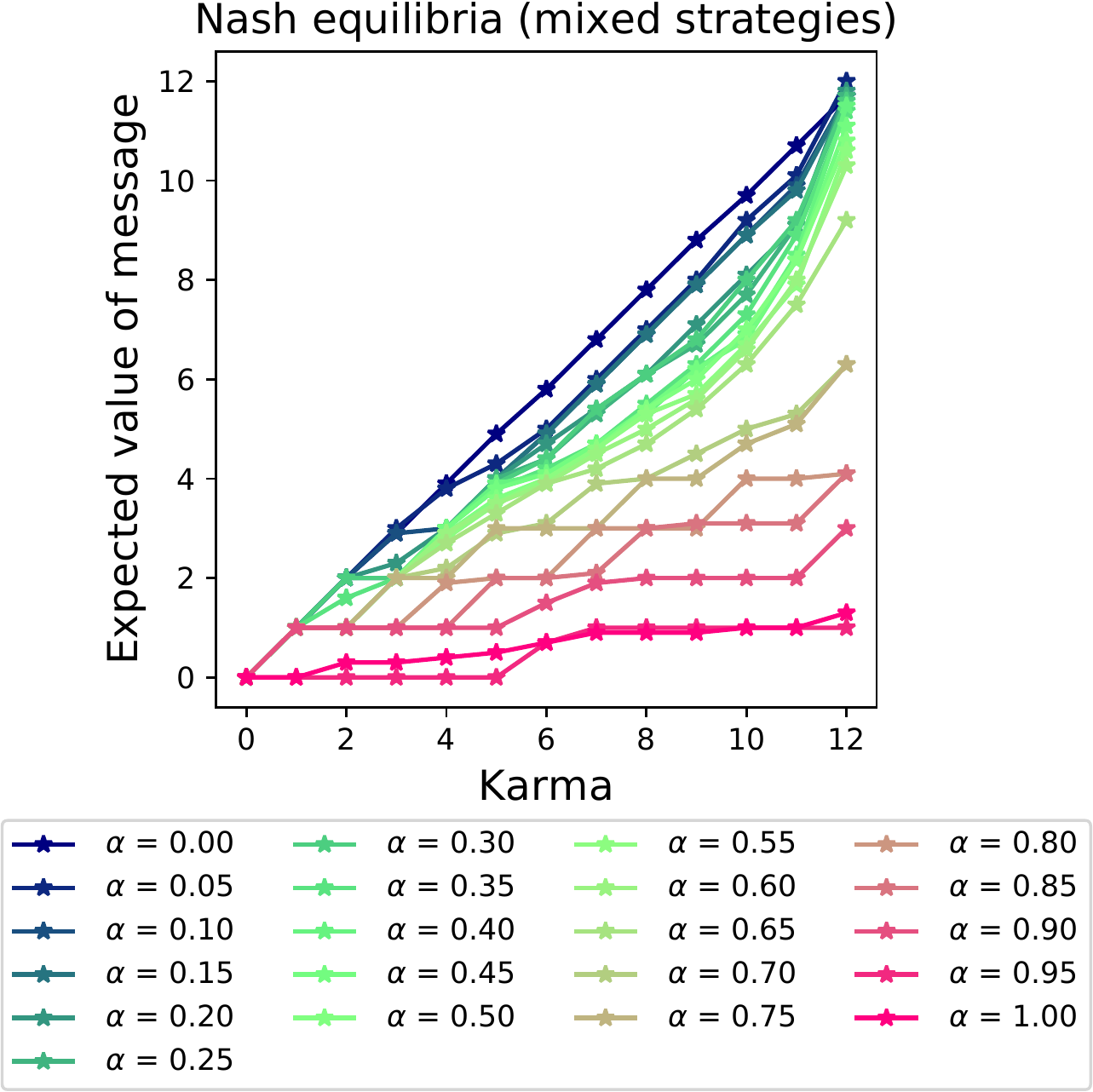}
    \caption{Expected message value given a karma level for mixed policies for different $\alpha$ discounting factors of future costs. Strategies with small discounting factor spend almost all available karma on a message whereas strategies with a large discounting factor save karma for the future.}
    \label{fig:nash_eq_overview}
\end{figure}

\subsection{Characterization of Nash equilibria for a Karma Game}

To characterize the equilibrium of the game, we must consider, in addition to the policy,
a series of other related quantities. These are:
\begin{itemize}    
\item $\stationary \in \PD(\PK)$ is the stationary distribution of karma values. Figure~\ref{fig:definitions-stationary} shows a typical stationary distribution.
\item $\transitions: \PK \to \PD(\PK)$ is a transition function for the karma levels; Figure~\ref{fig:definitions-transitions} shows a representation of such a transition function.
\item $\avgcost: \PK \to \realp$ is the expected cost of one interaction, as a function of the agent's karma.
\end{itemize}

The transition function $\transitions$ immediately descends from the composition of $\phi$ with the policy $\pi$ and the outcome distribution $\gamma$, assuming the  karma distribution~$D$ for the other agents and~$p$ for all agents' urgencies.

To express $\avgcost$, it is convenient to define the function $\rho(u_i,k_i,m_i)$ which gives the expected utility of choosing message $m_i$ for an agent of type $u_i,k_i$.
\newcommand{\var}[1]{{#1}}
\begin{multline}
\rho(u_i,k_i,m_i) =  
        \sum_{k_j \in \PK} \var{\stationary}_{k_j}
        \sum_{u_j \in \PS}  p_{u_j} \cdot \\
        \sum_{m_j \in \PM} \var{\policy}_{m_j}(u_j, k_j)
        \sum_{o \in \PO}
        \po_o(k_i, m_i, k_j, m_j) \cdot \\
        \left[
        c(o, u_i)  + \alpha \sum_{k'\in\PK} 
  \ft_{k'}(k_i, m_i, k_j, m_j, o)
  \var{\exputility}(k')
        \right].
\label{eq:rho}
\end{multline}

Figure~\ref{fig:definitions-utilities} shows a representation of a typical $\rho$.
Based on this definition, the expected cost of an interaction is
\begin{equation}
\avgcost(k_i) = 
\sum_{u_i \in \PS} p_{u_i}
\sum_{m_i \in \PM} \policy_{m_i}(u_i, k_i)
\rho(u_i,k_i,m_i).
\label{eq:expectedcost}
\end{equation}
 
We can now define the notion of Nash equilibrium for a Karma Game.

\begin{definition}[Nash equilibrium for a Karma Game]\label{def:nash_eq}
A policy $\policy$ is a Nash equilibrium for the Karma Game $G$ if 
there exist $\stationary, \transitions, \avgcost$ that satisfy three properties:

\textbf{P1: Stationarity}: $\stationary$ is the equilibrium distribution for the transition map $\transitions$:
\[
\stationary = \textstyle{\sum_{\tau \in \PK}}\, \stationary_\tau \transitions(\tau).
\]

\textbf{P2: Bellman}: There exists a function $\exputility: \PK \to \realp$, representing the expected total cost for an agent as a function of the present value of the karma, that satisfies the Bellman-like equation
\begin{equation} \label{eq:exputility-fixed-point}
    \exputility(k) = \avgcost(k) + \alpha \sum_{\tau \in \PK}
        \transitions_\tau(k) \exputility(\tau)
\end{equation}
for the expected interaction cost $\avgcost$ defined in \eqref{eq:expectedcost} and the discount factor $\alpha$.

\textbf{P3: Rationality}: The policy $\policy$ must yield the best expected outcome:
\[
C(\pi) \le C(\pi') \quad \forall \pi',
\]
where $C$ was defined in \eqref{eq:totalcost} and can be expressed as 
\[
C = \sum_{k_i \in \PK} \stationary_{k_i} \exputility(k_i).
\]
\end{definition}

The next section will be devoted to the numerical computation of a Nash equilibrium for the Karma Game of interest and to the interpretation of the resulting policies and outcome.

\section{Computing Nash equilibria of Karma Games}
\label{section:computing}

In general, Nash equilibria can be computed by iterative algorithms. Starting with an initial policy, one computes the other unknown (stationary distribution, karma utility); then one re-computes the optimal policy. If the recomputed policy is different from the initial one, the delta is a profitable perturbation of the policy. Based on the perturbation, one can make a small update of the policy, and repeat the process until convergence. If this process converges to a distribution, then by definition, we have found a Nash equilibrium as defined above. However, there is in general no guarantee that the iterative process converges.

\subsection{Fixed point computation} \label{sec:fixed}

We show here how to rearrange the equations to put them in the form of a fixed point.

Suppose we have a current guess of the policy $\policy$,
the stationary distribution $\stationary$, and the utility $\exputility$. 

\textbf{Step 1:} Compute the policy $\policy$ from the previous policy, the stationary distribution $\stationary$, and the expected utility $\exputility$.
The policy is computed using~\eqref{eq:policy-soft} based on the values of~$\rho$ obtained from~\eqref{eq:rho}.

\textbf{Step 2:} Compute the transitions~$\transitions$ from the policy~$\policy$ and the stationary distribution~$\stationary$.
Given the policy and the stationary distribution, we can compute the transitions of the system.
For each type~$u_i, k_i$, we know the distribution of the types it will encounter, and we know their policy. Thus, we can compute the outcomes, and the consequences of the outcomes in terms of what will be the next value of~$k_i$.

\textbf{Step 3:} Compute the stationary distribution $\stationary$ from the transitions $\transitions$.
This is a standard step - given a transition matrix, compute the equilibrium distribution. It can be done by iteration or by solving an eigenvector problem.

\textbf{Step 4:} Compute the expected utility $\exputility$ from $\stationary$ and $\policy$.
We can compute the expected utility using~\eqref{eq:exputility-fixed-point}. 
The expected daily cost $\overline{c}(k)$ is computed by setting $\alpha=0$ in~\eqref{eq:rho}.

\subsection{Momentum and simulated annealing}

We found  two simple devices that make the convergence robust, in the sense that the policy converges to the same solution no matter the initial conditions of the policy, stationary distribution, and karma utility.

\subsubsection{Momentum}

In Section~\ref{sec:fixed}, we have defined a way to update the policy $\policy$
that we can abstract as a function $\Psi$ such that:
$$
\policy_t^{\text{new}} = \Psi(\policy_t, \stationary_t, \exputility_t).
$$
Define the ``momentum'' $\tau$ as a scalar $0<\tau\leq 1$. Then we update the policy as
$$
\policy_{t+1} = \tau  \policy_t^{\text{new}}  +(1-\tau) \policy_{t}.
$$
For the set of simulations described below the optimization parameters were constant, but we did find in general that for different values of the model properties, the optimization parameters had to be optimized.

\subsubsection{Simulated annealing}

Let $T > 0$ be a temperature parameter.
Rather than looking  for a pure strategy, we set
\begin{equation}\label{eq:policy-soft}
\policy(u_i, k_i, m_i) \propto \exp(- \rho_{u_i,k_i}(m_i) / T).
\end{equation}
For large values of $T$, agents choose a random action. As $T$ decreases, the agents choose more often actions with good rewards.
As $T\to 0$,
the  policy tends to the deterministic policy,
where we select the maximum of $\rho_{u_i,k_i}(\cdot)$:
$$
\policy(u_i, k_i, m) \stackrel{T\to 0}{\longrightarrow}
\begin{cases}
1,& \text{if $m$ maximizes  $\rho_{u_i,k_i}(m)$}, \\
0,& \text{otherwise}.
\end{cases}
$$
In the simulations, we gradually decrease the temperature of the system in a series of ``eras''
(Figure~\ref{fig:policy_evolution}).

\subsection[Equilibria parametrization in alpha]{Equilibria parametrization in $\alpha$}

The parameter $\alpha$ introduced as a cost discounting factor in~\eqref{eq:totalcost} determines how much importance an agent assigns to future costs. 
In the limit $\alpha \ll 1$, the agent is only occupied with minimizing instantaneous costs. When $\alpha$ approaches 1, future costs are deemed almost as important as present costs.
To determine the influence this factor has on agent policies, we ran experiments with different $\alpha$ values ranging from $0$ to $1$ in $0.05$ increments. 
As an overview of the effect, we provide Figure~\ref{fig:discounting} which depicts the gradual changes in policy as $\alpha$ is increased. Similarly we offer Figure~\ref{fig:nash_eq_overview} as an overview of the effect of time discounting on the best message to send given a karma level.

One caveat that we have is that the Nash equilibria are not well defined when $\alpha = 1$ as some of the series in the formalization do not converge. Still, we also include the results of the algorithm for~$\alpha=1$. Similarly, we believe that for $\alpha \rightarrow 1$ there are numerical instabilities, and in fact we find that there are much larger oscillations. Rather than tuning the optimization parameters for each $\alpha$, we keep the same parameters, and we still picture the results for $\alpha=0.9$ and $\alpha=0.95$, without fully believing they are Nash equilibria for the game. 

\todo[inline]{Put here the figure with the optimized policies.}

\section{Policy comparison}

In this section, we are interested in gaining an empirical understanding of different solutions to the proposed distributed interaction problem. 

\paragraph{Evaluation protocol}

All simulations of interactions follow the same general procedure. As described in Section~\ref{section:model}, agents randomly meet in pairs and bid karma if they are urgent in order to pass first in an intersection. 
All experiments were conducted with 200 agents and a total of 1000 time periods. 
On each day, there are an average of 0.1 interactions per agent.
Agents are urgent with magnitude $3$ with probability $0.5$ and not urgent (magnitude $0$) again with probability $0.5$.
Each agent has an initial karma level uniformly randomly chosen between $0$ and $12$.
Agents can, through interactions, attain a minimum karma level of $0$ and a maximum karma level of $12$.

In the following, we compare various policies as well as the underlying parameters influencing the agents' policies.
We consider two performance metrics which are finite-sample proxies for $W_1$ and $W_2$, respectively:
\begin{itemize}
    \item ``Inefficiency'': This is the average cost per interaction attained by the agent
    at the end of the simulation period. Note that this is not the $\alpha$-discounted factor that 
    each agent is trying to minimize; rather, this is the social welfare---which roughly corresponds to the case $\alpha=1$.
    \item ``Unfairness'': This is the standard deviations of the costs at the end of the simulation period.
\end{itemize}

\begin{figure}[t]
    \centering
    \includegraphics[scale=0.7]{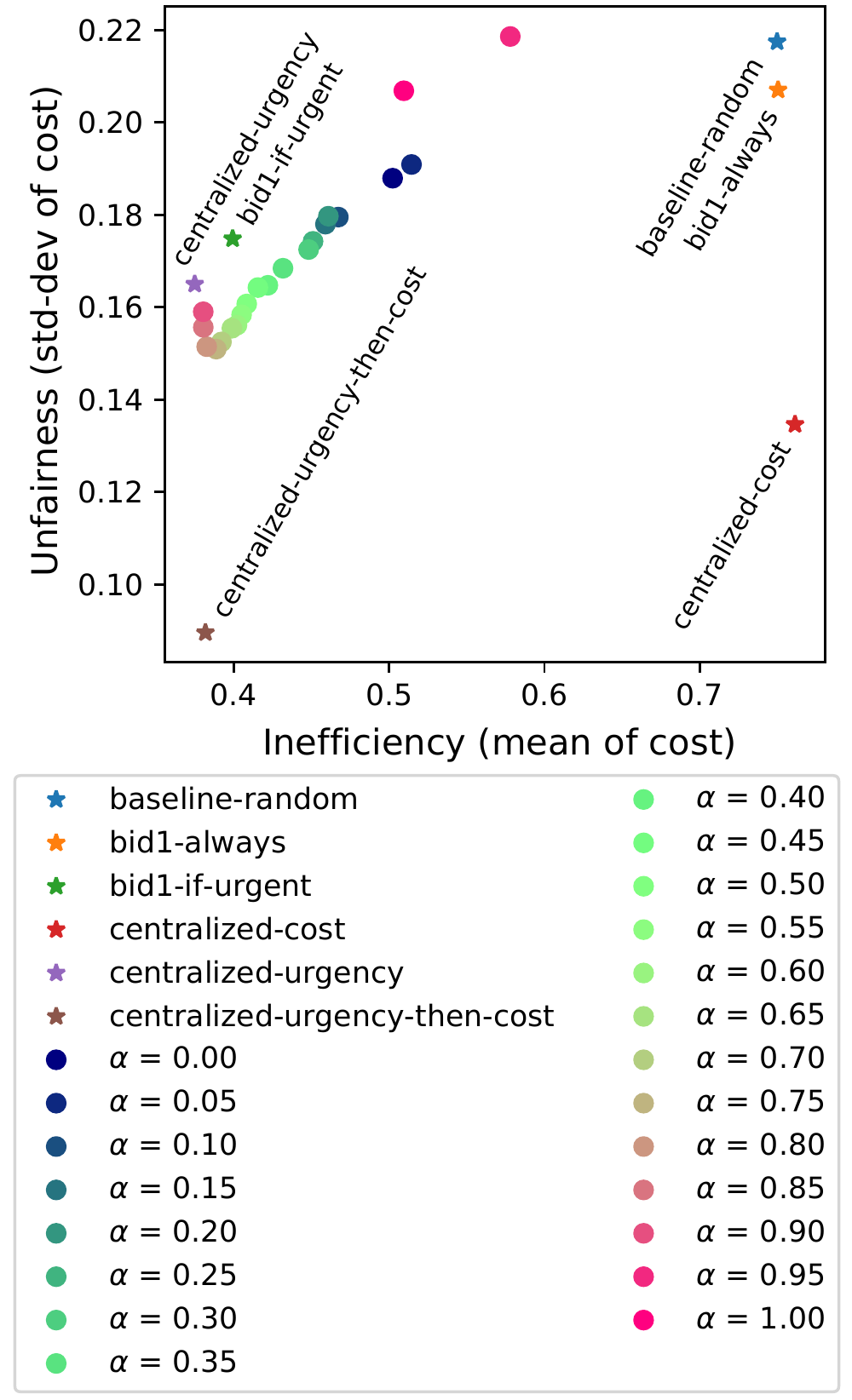}
    \caption{Overview of efficiency and unfairness of random, centralized and karma-based strategies. Random solutions fare the worst in both domains whereas centralized solutions with access to all information are optimal to their respective objectives. Karma-based solutions describe a trend of better efficiency and fairness with increasing discount factor $\alpha$ up to a limit.}
    \label{fig:results_overview}
\end{figure}

\paragraph{Policies}

In addition to the Nash equilibria found for sweeping $\alpha$  between 0 and 1,
we consider these other policies, as they are useful reference points:

% These are the policies used in the evaluation:

\newcommand{\code}[1]{\textsf{\small#1}}

\begin{itemize}
\item\code{baseline-random}: The winner is decided randomly.
\item\code{bid1-always}: The agents always bid 1.
\item\code{bid1-if-urgent}: The agents bid 1 if the urgency is nonzero, and zero otherwise.
\item\code{centralized-cost}: The policy  \eqref{eq:centralized-cost}.
\item\code{centralized-urgency}: The policy  \eqref{eq:centralized-urgency}.
\item\code{centralized-urgency-then-cost}: The policy  \eqref{eq:centralized-urgency-then-cost}.
\end{itemize}

\paragraph{Results}

The overall results are shown in Figure~\ref{fig:results_overview}. 

\code{baseline-random} (top right) obtains the worst results, as one might expect.

\code{centralized-urgency-then-cost} (bottom left) obtains the best results for both fairness and efficiency, as expected.

\code{centralized-cost} does well in terms of unfairness, as it tries to reduce the spread
    of the costs, but it is very inefficient.
    
\code{centralized-urgency} obtains minimum inefficiency (as predicted), but it does not do anything to reduce the spread of the costs, leading to a relatively high unfairness.

The baselines provide a reference frame to interpret the results for the karma-based policies.

We find many interesting nuggets. For example, \code{bid1-always} is very inefficient, as inefficient
as \code{baseline-random}, but it is less unfair. This is because the karma accounting keeps track of previous times when the agent lost, thereby slightly reducing the unfairness even if the policy is trivial.

Next consider the performance of \code{bid1-if-urgent}. This corresponds to a mechanism in which 
the agents use the karma message to reveal their urgency. Notwithstanding the fact that this is not an equilibrium for the game (this can be easily verified by noting that this is not a fixed point of the procedure discussed above), what we found surprising is that the efficiency is \emph{not} as good as some of the Nash equilibria that we find. 

Next we consider the performance of the Nash equilibrium as a function of $\alpha$. The sequence draws a hook in the inefficiency/unfairness space. The continuity of this curve also is good evidence that the procedure converged well (as noted before, for $\alpha\geq0.9$ the convergence is not assured).

We find the surprising result that for $\alpha\geq0.4$, the Nash equilibria are better in efficiency than the \code{bid1-if-urgent} strategy. The reason is that the agents should bid more or less if their karma levels allow---bidding only 1 is not the best strategy (neither for the agents nor society).
For $\alpha<0.4$, the agents do worse. 

The $\alpha=0$ ``there is no tomorrow'' strategy (bid everything if urgent) is particularly bad for society, though not as bad as random: karma still allows some reparations to be made.

We observe that for $\alpha>0.4$, the karma strategies beat the \code{centralized-urgency} strategy
in unfairness. There is a minimum unfairness observed for $\alpha=0.8$---we are not sure how this relates to the parameters of the problem. 

In these experiments, for $\alpha=0.85$, the performance is closest to the \code{centralized-urgency} strategy in both inefficiency and unfairness, in fact surprisingly close.

In conclusion, we obtain the surprising result that, for agents that are reasonably future-conscious, Nash equilibrium strategies beat heuristic solutions in both efficiency and fairness, and their performance is extremely close to the centralized solutions.

\section{Conclusions}
\label{section:conclusions}

We have demonstrated how the efficient use of a shared infrastructure can emerge from simple coordination protocols among competitive agents, without the need of any monetary transaction or complex decision infrastructures, in sharp contrast to most of the literature.
The enabler is the notion of \emph{karma}: a public state that links the decision of the same agent at different times (as long as each agent reasonably values its own future cost).
A solid understanding of the mechanisms that are necessary and sufficient for fair sharing of an infrastructure has the potential to guide the design of scalable solutions in many  applications, and in particular for autonomous mobility.

\addtolength{\textheight}{-12cm}   % This command serves to balance the column lengths

%%%%%%%%%%%%%%%%%%%%%%%%%%%%%%%%%%%%%%%%%%%%%%%%%%%%%%%%%%%%%%%%%%%%%%%%%%%%%%%%


%\section*{ACKNOWLEDGMENT}

%\section*{BIBLIOGRAPHY}

\bibliographystyle{IEEEtran}
\bibliography{karmagames}

\end{document}